\documentclass[aps,prb,reprint,superscriptaddress,showpacs,floatfix]{revtex4-1}

\usepackage[version=3]{mhchem} 
\usepackage{graphicx}
\usepackage{dcolumn}
\usepackage{bm}
\usepackage[mathlines]{lineno}
\usepackage{amssymb}
\usepackage{amsmath}

\usepackage{threeparttable}  
\usepackage{multirow}
\usepackage{booktabs}
\usepackage{color}
\usepackage{ulem}
\usepackage{soul}
\usepackage{textcomp} 
\definecolor{LightBlu}{RGB}{19, 183, 219}
\definecolor{GreyBlu}{RGB}{136, 177, 208}
\definecolor{GreyPur}{RGB}{132, 127, 197}
\definecolor{GreyRed}{RGB}{207, 111, 96}
\definecolor{GreyOra}{RGB}{230, 161, 86}
\usepackage{datetime}
\usepackage[colorlinks,linkcolor=blue,anchorcolor=blue,citecolor=blue,urlcolor=blue]{hyperref}

\begin{document}
\title{Role of octahedral tilting induced acoustic softening on limiting thermal transport in SrSnO$_3$}

\author{Yuzhou Hao}
\affiliation{State Key Laboratory for Mechanical Behavior of Materials, State Key Laboratory of Porous Metal Materials, 
School of Materials Science and Engineering, Xi’an Jiaotong University, Xi’an 710049, China}

\author{Turab Lookman}
\affiliation{AiMaterials Research LLC, Santa Fe, New Mexico 87501, USA}

\author{Xiangdong Ding}
\email{dingxd@mail.xjtu.edu.cn}
\affiliation{State Key Laboratory for Mechanical Behavior of Materials, State Key Laboratory of Porous Metal Materials, 
School of Materials Science and Engineering, Xi’an Jiaotong University, Xi’an 710049, China}

\author{Jun Sun}
\affiliation{State Key Laboratory for Mechanical Behavior of Materials, State Key Laboratory of Porous Metal Materials, 
School of Materials Science and Engineering, Xi’an Jiaotong University, Xi’an 710049, China}

\author{Zhibin Gao}
\email{zhibin.gao@xjtu.edu.cn}
\affiliation{State Key Laboratory for Mechanical Behavior of Materials, State Key Laboratory of Porous Metal Materials, 
School of Materials Science and Engineering, Xi’an Jiaotong University, Xi’an 710049, China}

\date{\today}



\begin{abstract}

{Octahedral tilting is a fundamental structural distortion in perovskites, governing key phenomena such as lattice stabilizing, soft phonon dynamics, group-theoretical analysis, phase transitions, ferroelectricity, and even for tunable electronic band gap. However, its influence on lattice thermal conductivity ($\kappa_L$) remains poorly understood. In the archetypal perovskite SrTiO$_3$, tilting in the low-temperature tetragonal phase is known to enhance $\kappa_L$ by suppressing specific phonon scattering channels around 200 cm$^{-1}$. Here, we investigate the thermal transport in strontium stannate (SrSnO$_3$), another perovskite oxide that undergoes temperature-driven phase transitions, and reveal a completely opposite effect. Through a systematic study across its orthorhombic, tetragonal, and cubic phases, we demonstrate that octahedral tilting in the tetragonal phase of SrSnO$_3$ anomalously triggers acoustic phonon softening. This softening manifests as reduced frequencies and group velocities in low-frequency (\textless 3 THz) acoustic modes, creating a large decrease for heat transport, particularly along the $c$-axis. Consequently, $\kappa_L$ is significantly suppressed, decreasing from 7.48 W m$^{-1}$ K$^{-1}$ to 6.06 W m$^{-1}$ K$^{-1}$ as the tilting angle increases by a mere 1$^\circ$. These findings identify tilting-induced acoustic softening as a pivotal mechanism for limiting and controlling anisotropic thermal transport in SrSnO$_3$, presenting a stark contrast to the established behavior in SrTiO$_3$.}
\end{abstract}

\maketitle

\section{Introduction}\label{sec1}
Perovskite oxides (ABO$_3$) exemplify the tunability of material properties, exhibiting functionalities such as ferroelectricity and superconductivity alongside adjustable thermal transport~\cite{martin2017perovskite,howard1998group,sim2013octahedral,iqbal2024composition,prasanna2017band}.
SrSnO$_3$ (SSO) is notable for its temperature-driven phase transitions: orthorhombic phase ($Pnma$) $<905$~K, high temperature orthorhombic phase (905--1062~K), tetragonal phase ($I4/mcm$) (1062--1295~K), and cubic phase ($Pm\overline{3}m$) $>1295$~K~\cite{angel2005general,shao2022unlocking,glerup2005high,singh2009structural,SupplementalMaterial}. These transitions stem from cooperative rotations (tilting) of SnO$_6$ octahedra, profoundly influencing lattice dynamics and thermal conductivity. While the bulk tetragonal phase of SSO is thermodynamically stable in the high-temperature range of 1062--1295~K, it can indeed be stabilized at lower temperatures (including room temperature) through epitaxial strain engineering in thin films. Experimental studies have successfully synthesized tetragonal-like SSO films on substrates such as GdScO$_3$ by utilizing lattice mismatch and strain engineering~\cite{Wang2018Engineering, Ghosh2025Defect}.

As a wide-bandgap semiconductor (approximately 4.1 eV), SSO holds promise for applications in transparent conductors and power electronics, where efficient thermal management is essential to prevent overheating and maintain performance under high-power operation~\cite{kim2022deep,zhang2023temperature}. Recent experimental studies report a room-temperature thermal conductivity of 4.4 W m$^{-1}$ K$^{-1}$ for SSO thin films, significantly lower than other perovskites like SrTiO$_3$ or BaSnO$_3$ (by about 60\%), suggesting a strong influence from octahedral tilting in its orthorhombic phase~\cite{zhang2023temperature}. Despite this, the thermal transport properties of SSO, particularly in its tetragonal phase with its unique symmetry and phonon characteristics, remain underexplored compared to well-studied perovskites like SrTiO$_3$ (STO)~\cite{fumega2020understanding,menahem2021strongly,lanigan2021two}.

Thermal conductivity is a critical material property governing heat management in diverse technologies, from microelectronics demanding efficient dissipation to thermoelectrics requiring low $\kappa_L$ for high conversion efficiency (quantified by $ZT$)~\cite{Snyder2008Complex,cahill2014nanoscale}. Tailoring $\kappa_L$ through structural design is thus a central challenge in materials science, particularly for complex oxides offering rich avenues for phonon engineering. In perovskites, octahedral tilting fundamentally alters phonon dispersion and scattering~\cite{howard1998octahedral}. This distortion arises from the rotation of SnO$_6$ octahedra around crystallographic axes, classified using a systematic notation that reflects the tilt pattern~\cite{howard1998octahedral}. 

In STO, tilting in the low-temperature tetragonal phase has been shown to increase $\kappa_L$ by suppressing specific scattering channels around 200 cm$^{-1}$, an effect attributed to shifts in phonon dispersion~\cite{fumega2020understanding}. 
Anomalously, previous study of Cs$_2$AgBiBr$_6$ showed that an increase in the octahedral tilting angle enhances anharmonicity, which in turn reduces the $\kappa_L$~\cite{Zheng2024Unravelling}. While SSO’s stannate framework exhibiting greater anharmonicity compared with STO due to the larger Sn$^{4+}$ ionic radius~\cite{Cohen1992Origin} (Sn$^{4+}$: 0.690 \AA~and Ti$^{4+}$: 0.605 \AA~\cite{Shannon1976Revised}), this could lead to differences in chemical bonding and lattice dynamics between SSO and STO~\cite{Tadano2015consistent}. The precise coupling between tilting, acoustic softening, and suppressed $\kappa_L$ in tetragonal SSO thus warrants detailed investigation beyond the work of $\kappa_L$ in the orthorhombic phase~\cite{zhang2023temperature}.

Accurate prediction of $\kappa_L$ requires accounting for not only three-phonon (3ph) scattering but also non-negligible four-phonon (4ph) processes and coherent phonon contributions ($\kappa_c$), which collectively necessitate extensive computational resources to capture complex anharmonic interactions via first-principles density functional theory (DFT) combined with solutions to the Boltzmann transport equation (BTE)~\cite{Zheng2022Anharmonicity,Zheng2022Effects,Xia2020Particlelike,Xia2020Microscopic,Zhao2021Lattice}. Probing these mechanisms computationally is highly demanding. However, directly computing the required high-order interatomic force constants (IFCs) in large supercells or across phase transitions—essential for materials like SSO with pronounced tilting—remains prohibitively expensive using conventional DFT~\cite{gao2018unusually,FENG2024anharmonicity,Wang2024thermoelectricity,Wang2024Revisiting}. This challenge is further amplified by the necessity to incorporate temperature-dependent phonon renormalization and the subtle impacts of structural distortions on low-frequency acoustic modes, both of which are pivotal for elucidating the observed $\kappa_L$ reduction.

This work provides first-principles insights into the origin of reduced $\kappa_L$ with increasing tilting angle in tetragonal SrSnO$_3$, establishing the critical role of tilting-enhanced acoustic phonon softening. Utilizing DFT and NEP-assisted computations, we analyze phonon properties and $\kappa_L$ anisotropy across orthorhombic, tetragonal, and cubic phases from 100~K to 1600~K. Our results demonstrate that octahedral tilting in the $I4/mcm$ phase induces significant acoustic mode softening, elevating scattering rates and suppressing $\kappa_L$ compared to the cubic phase. 
These findings advance the fundamental understanding of thermal transport in tilting perovskites.

\section{Methods}\label{sec3}
Machine Learning Potentials (MLPs) utilize machine learning algorithms to model the atomic potential energy surface, predicting both energy and forces based on the atomic environment, thus enabling DFT-level accuracy at reduced computational cost~\cite{Hao2024Machine,Ouyang2022Accurate,Bohayra2021Accelerating,behler2007generalized}. MLPs can be integrated with the Boltzmann Transport Equation (BTE) method to offer valuable insights into phonon transport mechanisms in complex systems. Among MLPs, the Neuroevolution Potential (NEP) has demonstrated an effective balance between computational efficiency and accuracy~\cite{Fan2022GPUMD,Dong2024Molecular}. NEP can produce precise second- and third-order force constants, and previous studies have shown its ability to reliably compute fourth-order force constants, which are critical for accurately predicting $\kappa_L$ in systems dominated by 4ph scattering~\cite{Ouyang2022Accurate}. NEP has also demonstrated success in modeling thermal conductivity across diverse materials, including metals, oxides, and liquids~\cite{Cao2025Lattice,xu2023accurate,timalsina2024neuroevolution}. 

Building on this foundation, we have refined and enhanced the NEP framework to efficiently calculate the fourth-order force constants, enabling more accurate modeling of thermal transport in materials with strong anharmonic effects~\cite{Fan2019Homogeneous,Denis1982NEMD,Gabourie2021Spectral}. The subsequent sections present the calculation methods employed in this work and the associated parameters.

The Vienna Ab initio Simulation Package (VASP) was used to optimize the crystal structure of SrSnO$_3$~\cite{SupplementalMaterial}. All DFT calculations were performed using the Projector Augmented Wave (PAW) method with the PBEsol exchange-correlation functional~\cite{Perdew1996Generalized,Perdew2008Restoring,Csonka2009Assessing}. A cutoff energy of 500 eV was applied for the wavefunction with force and energy convergence thresholds set to $10^{-3}$ eV/\AA\ and $10^{-8}$ eV respectively, for both structural relaxation and self-consistent calculations. The structural optimization process involved full relaxation of the cell and atomic positions, utilizing a Gamma-centered k-point grid of $6 \times 6 \times 4$. The PBEsol was selected because it is widely recognized to improve upon PBE for predicting the equilibrium lattice constants and structural distortions of solids, particularly perovskite oxides.
Our calculated equilibrium lattice constants of tetragonal SSO using PBEsol deviate by 2.7\% in $a$- and $b$-axes, 0.023\% in $c$-axis from experimental values reported~\cite{Fengdeng2024Deep, BEURMANN2003392}, and the calculated equilibrium tilt angle of $12.998^\circ$ is in agreement with the experimental value of $10^\circ$ at $0^\circ C$~\cite{Mountstevens2005Order}.

\textit{Ab initio} molecular dynamics (AIMD) simulations were performed to collect energy and atomic forces as training datasets. The isobaric-isothermal (NPT) ensemble, with a sampling time interval of 100 fs (AIMD integration time step 5 fs, sampling 1 out of every 20 steps), was applied to a $3 \times 3 \times 3$ primitive cell of tetragonal SrSnO$_3$, a $3 \times 3 \times 3$ primitive cell of cubic SrSnO$_3$, and a $2 \times 2 \times 2$ primitive cell of orthorhombic SrSnO$_3$, generating a total of 180 structures at temperatures ranging from 5 K to 350 K. A $3 \times 3 \times 3$ k-point grid was used. Additional datasets were created using random perturbations generated with the dpdata package~\cite{Deep2018Zhang,WANG2018DeePMD}. The unit cell deformation ratio was set to 5\%, with diagonal perturbations uniformly distributed within the range [$-5\%$, $5\%$]. The maximum atomic displacement radius was set to 0.25 \AA, meaning atomic positions were randomly and uniformly distributed within a spherical region of radius 0.25 \AA. This process generated 120 structures, with additional 120 different tilting angle structures of tetragonal SrSnO$_3$, resulting in a total training set of 420 structures. The validation dataset, containing 80 structures, was also generated using random perturbations package~\cite{Deep2018Zhang,WANG2018DeePMD}. 
The specific focus on 5 K-350 K AIMD data is tailored for the BTE method which requires high-precision interatomic force constants derived from finite displacement theory which calculates structures near the equilibrium position. The low-temperature AIMD data provides a dense sampling of these near-equilibrium forces, ensuring high accuracy for the harmonic and anharmonic force constants, while the perturbed structures ensure robustness against structural instability at high temperatures.

To model interatomic interactions, NEP were trained on datasets derived from DFT calculations~\cite{Fan2022GPUMD,Dong2024Molecular}. The NEP parameters were optimized by minimizing the following loss function,
\begin{equation}
\begin{aligned}
L(z) =& \lambda_e \left( \frac{1}{N_{\text{str}}} \sum_{n=1}^{N_{\text{str}}} \left( U^{\text{NEP}}(n, z) - U^{\text{tar}}(n) \right)^2 \right)^{1/2} \\
& + \lambda_f \left( \frac{1}{3N} \sum_{i=1}^{N} \left( F_i^{\text{NEP}}(z) - F_i^{\text{tar}} \right)^2 \right)^{1/2} \\
& + \lambda_v \left( \frac{1}{6N_{\text{str}}} \sum_{n=1}^{N_{\text{str}}} \sum_{\mu\nu} \left( W_{\mu\nu}^{\text{NEP}}(n, z) - W_{\mu\nu}^{\text{tar}}(n) \right)^2 \right)^{1/2} \\
& + \lambda_1 \frac{1}{N_{\text{par}}} \sum_{n=1}^{N_{\text{par}}} |z_n|\\
& + \lambda_2 \left( \frac{1}{N_{\text{par}}} \sum_{n=1}^{N_{\text{par}}} z_n^2 \right)^{1/2},
\end{aligned}
\end{equation}
where $N_{\text{str}}$ represents the number of structures in the training dataset, and $N$ is the total number of atoms in these structures. The first three terms measure the root-mean-square errors (RMSE) of energy, forces, and virial tensors predicted by NEP compared to DFT values. The last two terms incorporate $\ell_1$ and $\ell_2$ regularization. The weights$\lambda_e$, $\lambda_f$, $\lambda_v$, $\lambda_1$, and $\lambda_2$ determine the relative weights of these terms. Energy and virial tensors are expressed in units of eV/atom, while force components are expressed in eV/\AA. The accuracy of NEP are specified in Fig. S2 (the plot of energy per atom, forces, stresses and phonon dispersion curves predicted by the NEP model versus DFT calculations), and Fig. S8 (the plot of energy per atom, forces predicted by the NEP model versus DFT calculations at high temperature over 1400 K in cubic SSO).

The NEP fitting cutoff radius for radial and angular descriptors was set to 7.0 \AA, exceeding the interaction cutoff values used for cubic and quartic phonon calculations (4.3 \AA\ and 3.0 \AA, respectively), ensuring sufficient accuracy for thermal conductivity modeling. To further enhance precision, the number of radial and angular basis functions was increased to 14, higher than the default value of 8.

The lattice thermal conductivity $\kappa_p$, based on the Peierls-Boltzmann transport theory, is expressed as~\cite{WuLI2014ShengBTE, Han2022FourPhonon},
\begin{equation}
\kappa_p = \frac{\hbar^2}{k_B T^2 V N_0} \sum_{\lambda} n_\lambda (n_\lambda + 1) \omega^2_\lambda \bm{v}\lambda \otimes \bm{v}\lambda \bm{\tau}_\lambda,
\end{equation}
where $\hbar$ denotes the reduced Planck constant, $k_B$ the Boltzmann constant, $T$ the absolute temperature, $V$ the volume of the primitive unit cell, and $N_0$ the number of phonon wave vectors sampled in the first Brillouin zone. The variables $n_\lambda$, $\omega_\lambda$, $\bm{v}_\lambda$, and $\tau_\lambda$ represent the phonon population, frequency, group velocity, and lifetime for the $l$ mode, which is defined by wave vector $q$ and branch index $s$. The coherent term $\kappa_c$, accounting for off-diagonal heat-flux contributions between phonon modes, is defined as~\cite{Simoncelli2019},

\begin{equation}
\begin{aligned}
\kappa_{c}= & \frac{\hbar^{2}}{k_{B} T^{2} V N_{0}} \sum_{\boldsymbol{q}} \sum_{s \neq s^{\prime}} \frac{\omega_{\boldsymbol{q}}^{s}+\omega_{\boldsymbol{q}}^{s^{\prime}}}{2} \boldsymbol{v}_{\boldsymbol{q}}^{s, \boldsymbol{s}^{\prime}} \boldsymbol{v}_{\boldsymbol{q}}^{\boldsymbol{s}^{\prime}, s} \\
& \times \frac{\omega_{\boldsymbol{q}}^{s} n_{\boldsymbol{q}}^{s}\left(n_{\boldsymbol{q}}^{s}+1\right)+\omega_{\boldsymbol{q}}^{s^{\prime}} n_{\boldsymbol{q}}^{s^{\prime}}\left(n_{\boldsymbol{q}}^{s^{\prime}}+1\right)}{4\left(\omega_{\boldsymbol{q}}^{s^{\prime}}-\omega_{\boldsymbol{q}}^{s}\right)^{2}+\left(\Gamma_{\boldsymbol{q}}^{s}+\Gamma_{\boldsymbol{q}}^{s^{\prime}}\right)^{2}} \\
& \times\left(\Gamma_{\boldsymbol{q}}^{s}+\Gamma_{\boldsymbol{q}}^{s^{\prime}}\right),
\end{aligned}
\end{equation}
where $\boldsymbol{v}_{\boldsymbol{q}}^{s, \boldsymbol{s}^{\prime}}$ represents the off-diagonal group velocity coupling modes $s$ and $s'$, and $\Gamma_{\boldsymbol{q}}^{s}$ denotes the phonon linewidth. Finally, $\kappa_{c}$ is calculated by our homegrown code based on Eq. (3).

For the interatomic force constants (IFCs) calculations, a $3 \times 3 \times 2$ supercell of SrSnO$_3$ was used. Third- and fourth-order IFCs were computed up to the fifth and second nearest neighbors, respectively, using a $5 \times 5 \times 5$ q-mesh for four-phonon processes. The q-mesh was also verified to be converged, as shown in Fig. S7(c). Phonopy, thirdorder.py, fourthorder.py together with our custom-built interface for NEP are used to calculate the second-, third- and fourth-order IFCs~\cite{TOGO2015First,WuLI2014ShengBTE,Feng2017Four,Han2022FourPhonon}. The Self-Consistent Phonon (SCPH) approximation was employed to correct phonon frequencies for quartic anharmonicity~\cite{Xia2020Particlelike,Xia2020Microscopic}, which is especially critical for materials with soft phonon modes. The SCPH approach in CSLD~\cite{Xia2020Throughput, Zhou2014Lattice} is described as~\cite{Xia2020Particlelike, Xia2020Microscopic}:

\begin{equation}
{\Omega}_{\lambda}^{2} = {\omega}_{\lambda}^2+2{\Omega}_{\lambda}\sum\limits_{{\lambda}_{1}} I_{\lambda\lambda_1},
\end{equation}
where $\omega_\lambda$ represents the initial phonon frequency derived from the harmonic approximation, while ${\Omega}_{\lambda}^{2}$ denotes the temperature-dependent renormalized phonon frequency. The scalar $I_{\lambda\lambda_1}$ is calculated as follows~\cite{Xia2020Particlelike,Xia2020Microscopic},
\begin{equation}
{I_{\lambda\lambda_1}} = \frac{\hbar}{8 N_0} \frac{V^{(4)} (\lambda,-\lambda,\lambda_1,-\lambda_1)}{\Omega_{\lambda}\Omega_{\lambda_1}} \left[1 + 2n_\lambda(\Omega_{\lambda_1})\right],
\end{equation}
where $V^{(4)}$ represents the fourth-order IFCs in the reciprocal space, and the phonon population $n_\lambda$ follows the Bose-Einstein distribution. Both equations include the parameters $I_{\lambda\lambda_1}$ and $\Omega_\lambda$, enabling the SCPH equation to be solved iteratively.

The suppressed $\kappa_L$ due to phonon-boundary scattering is expressed as~\cite{Negi2023Thickness},
\begin{equation}
\kappa_{nan}=\int_{0}^{\infty}\kappa_{bulk}(\lambda)\cdot B\left(\frac{\lambda}{d}\right)d\lambda
\end{equation}
where $B(\lambda/d)$ is the suppression function. For the through-plane thermal conductivity in thin films, B is approximated by,
\begin{equation}
B_{film}(Kn) = 1+3Kn\left[E_5(Kn^{-1}) - 0.25\right]
\end{equation}
with $Kn=\lambda/d$ and $E_5$ is the 5th-order exponential integral.

Vacancies reduce $\kappa_L$ by introducing additional phonon scattering. The vacancy scattering rate is calculated as~\cite{Tiwari2023Intrinsic},
\begin{equation}
\tau_{vacancy}^{-1}=\frac{9\pi}{2}f_{v}\omega^{2}pDOS(\omega)
\end{equation}
where $f_v$ is the vacancy concentration, $\omega$ is the phonon frequency, and $pDOS(\omega)$ is the partial density of states of the atom species associated with the vacancy. The constant 9 accounts for mass and bond-disorder effects.

\section{Results}\label{sec2}


\begin{figure*}
\centering
\includegraphics[width=0.9\textwidth]{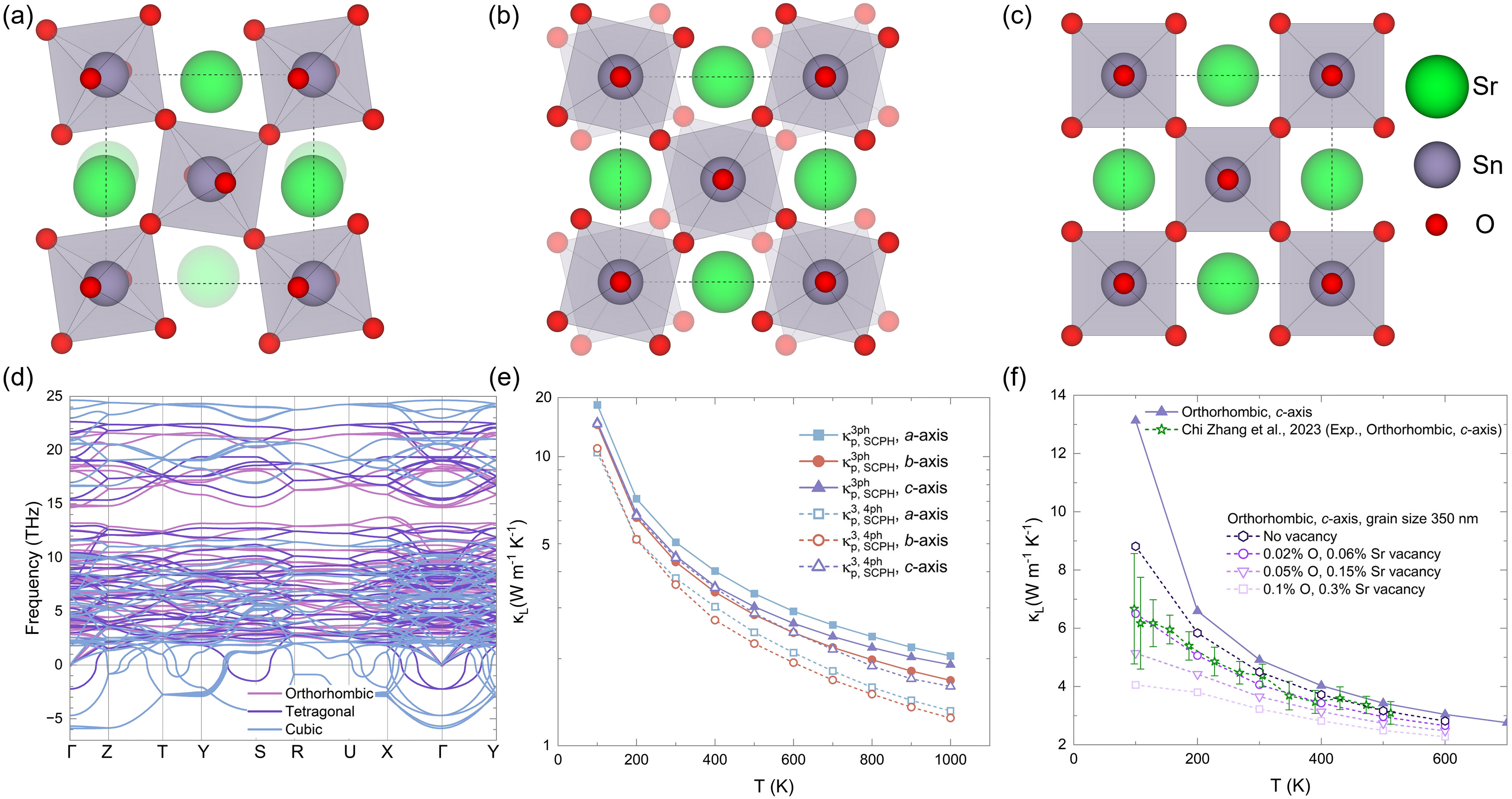}
\caption{Crystal structure of (a) orthorhombic phase, (b) tetragonal phase, and (c) cubic phase of SrSnO$_3$ viewed along the $c$-axis, showing the arrangement of Sr (green), Sn (gray), and O (red) atoms in a perovskite lattice. (d) Phonon dispersion curves of the orthorhombic, tetragonal phase, and cubic phase of SrSnO$_3$, with high-symmetry points labeled along the Brillouin zone path ($\Gamma$--Z--T--Y--S--R--U--X--$\Gamma$--Y). (e) Lattice thermal conductivity with 3ph and 4ph interactions as a function of temperature. (f) Comparison of the c-axis thermal conductivity of the orthorhombic phase with experimental values considering grain boundary scattering and vacancy scattering~\cite{zhang2023temperature}.}
\label{fig1}
\end{figure*}


Fig.~\ref{fig1}(a) displays the orthorhombic phase of SrSnO$_3$ (space group $Pnma$, $T$ $<$ 905 K) viewed along the $c$-axis, showcasing a distorted perovskite structure characterized by significant tilting of SnO$_6$ octahedra. The atomic arrangement features Sr (green) at the A-site, Sn (grey) at the B-site forming octahedra, and O (red) at vertices. This tilting is a hallmark of lower-symmetry perovskites arising from cooperative octahedral rotations, reducing lattice symmetry and introducing anisotropy. The distortions enhance phonon scattering through altered bond lengths and angles, consistent with observations in perovskites like SrTiO$_3$ where octahedral tilting modulates thermal transport \cite{fumega2020understanding}. 

\begin{figure*}
\centering
\includegraphics[width=0.9\textwidth]{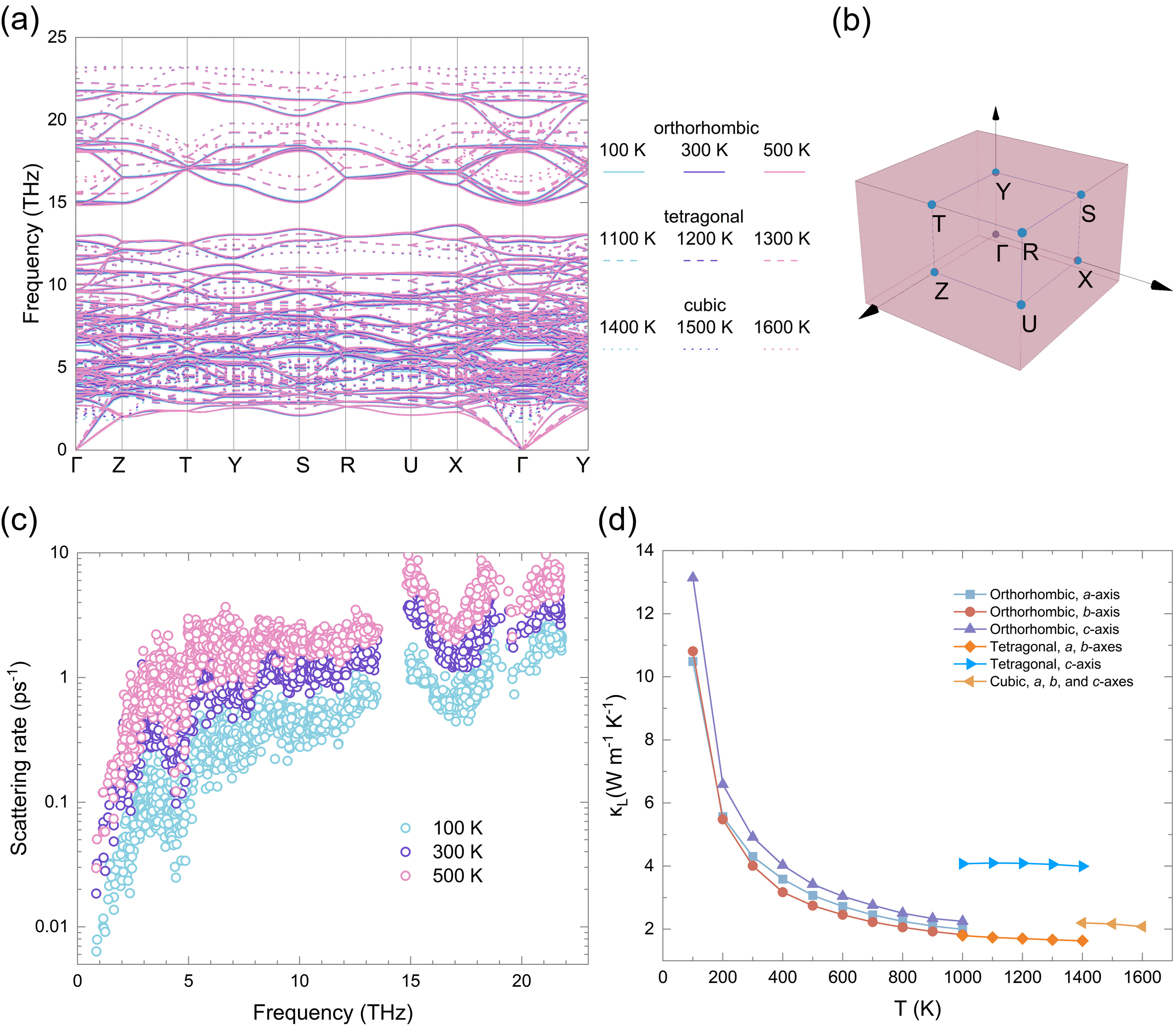}
\caption{(a) Renormalized phonon dispersion curves of all three SSO phases (orthorhombic, tetragonal, and cubic) as a function of temperature, with high-symmetry points labeled along the Brillouin zone path ($\Gamma$--Z--T--Y--S--R--U--X--$\Gamma$--Y). The temperatures ranging from 100 K to 500 K for orthorhombic phase, 1100 K to 1300 K for tetragonal phase, and 1400 K to 1600 K for cubic phase. (b) Brillouin zone path for the phonon dispersion. (c) Phonon scattering rate of the orthorhombic phase as a function of temperature from 100 K to 500 K. (d) Lattice thermal conductivity ($\kappa_L$) of SrSnO$_3$ as a function of temperature (100--1600 K).
\label{fig2}}
\end{figure*}

Fig.~\ref{fig1}(b) illustrates the tetragonal phase (space group $I4/mcm$, 1062 $>$ $T$ $>$ 1295 K) viewed along the $c$-axis, exhibiting reduced octahedral tilting compared to the orthorhombic phase. SnO$_6$ octahedra are less distorted, with Sr atoms at unit cell corners and O atoms forming an open framework along $c$. This structural relaxation correlates with changes in phonon group velocities and scattering rates shown in Fig.~\ref{fig2}, potentially enhancing thermal conductivity along $c$-axis, as explored in the $\kappa_L$ analysis. Fig.~\ref{fig1}(c) depicts the cubic phase (space group $Pm\bar{3}m$, $T>$ 1295 K) viewed along the $c$-axis, representing the high-symmetry state with no  tilting. Perfectly aligned SnO$_6$ octahedra and symmetric atomic placements minimize structural anharmonicity, yielding isotropic thermal conductivity~\cite{wang2023role}.

Fig.~\ref{fig1}(d) shows phonon dispersion of all three phases along $\Gamma$-Z-T-Y-S-R-U-X-$\Gamma$-Y at 0 K. The tetragonal and cubic phases exhibit significant imaginary frequencies in their phonon spectra, whereas the orthorhombic phase shows none imaginary frequencies, indicating its stability at low temperatures. Optical modes (from $\sim\!3\,\text{THz}$ at $\Gamma$) interact with acoustic modes, suggesting strong anharmonic coupling that suppresses thermal transport. 

Fig.~\ref{fig1}(e) shows that in the low-temperature orthorhombic phase, $\kappa_L$ including 4ph scattering is significantly lower than that with only 3ph scattering in all three directions. At 300 K, $\kappa_{p, SCPH}^{3ph}$ in three directions are 5.05, 4.32, and 4.53 $\text{W m}^{-1}\text{ K}^{-1}$, while $\kappa_{p, SCPH}^{3, 4ph}$ are 3.80, 3.61, and 4.46 $\text{W m}^{-1}\text{ K}^{-1}$, representing reductions of 25\%, 16\%, and 2\%, respectively. At 1000 K, these reductions become 35\%, 26\%, and 16\%, respectively. This indicates that the 4ph effect plays a non-negligible role in $\kappa_L$ of SSO, and its importance increases with rising temperature. 

Therefore, to obtain accurate information on $\kappa_L$ of SSO, the 4ph effect must be considered. In Fig.~\ref{fig1}(f), we have taken the 4ph effect into account and also included the coherent phonons, $\kappa_{c}$.
Boundary effects (see Methods Section) reduce $\kappa_L$ in orthorhombic phase along $c$-axis with 350\,nm grain size which is consistent with sample thickness in reference~\cite{zhang2023temperature} (8.82 $\text{W m}^{-1}\text{ K}^{-1}$ at 100 K and 4.50 $\text{W m}^{-1}\text{ K}^{-1}$ at 300 K), showing a 33\% and 8\% drop versus bulk at 100 K and 300 K. Since the exact vacancy concentrations were not quantified in the experiment, we treated these as variable parameters calibrated within a physically reasonable range for MBE-grown oxide films. Sr (0.06\%, 0.15\%, and 0.3\%) and O (0.02\%, 0.05\%, and 0.1\%) vacancies further suppress $\kappa_L$ via point-defect scattering, with higher concentrations amplifying reductions. Experimental $\kappa_L$ in $c$-axis for 350\,nm orthorhombic films \cite{zhang2023temperature} green stars, $\sim\!6.67\,\text{W m}^{-1}\text{ K}^{-1}$ at 98\,K, $\sim\!4.37\,\text{W m}^{-1}\text{ K}^{-1}$ at 305\,K to $\sim\!3.09\,\text{W m}^{-1}\text{ K}^{-1}$ at 512 K) align with simulations incorporating boundary and 0.06\% Sr, 0.02\% O vacancy defect effects (5\%, 8\%, and 4\% error), validating computational approaches assisted by accurate machine learning potential.

Fourth-order anharmonicity not only induces a shift in phonon frequencies but also produces significant 4ph scattering~\cite{feng2024relation, Xia2020Throughput, Wang2024thermoelectricity}. To delve deeper into the temperature effects, we separately investigated the influence of temperature on the temperature-dependent phonon spectrum in Fig.~\ref{fig2}(a) and the phonon scattering rates in Fig.~\ref{fig2}(c). Fig.~\ref{fig2}(a) displays the temperature-dependent phonon spectra corresponding to the renormalized second-order force constants for the three phases. It is evident that there are significant changes in the phonon spectra among the three phases due to the change in symmetry. However, within a single phase the temperature effect is relatively weak, with only slight phonon hardening observable around 5 THz. In contrast, Fig.~\ref{fig2}(c) shows the variation of phonon scattering rates (orthorhombic phase) with temperature. Scattering rates for higher temperature ranges of SSO are also shown in Fig. S4. It can be seen that 4ph scattering rates rise with temperature, and lead to SSO achieving a low thermal conductivity below 5 $\text{W m}^{-1}\text{ K}^{-1}$ in the mid-to-high temperature range.

In Fig.~\ref{fig2}(d), we present the temperature dependence of $\kappa_L$ in each direction for the three phases, including $\kappa_{p}$ and $\kappa_{c}$. In the calculation of $\kappa_{p}$, both the effect of phonon frequency shifts and the 4ph scattering effect are considered.
In the orthorhombic phase, $\kappa_L$ exhibits pronounced anisotropy: values along $c$-axis (purple, 13.13 $\text{W m}^{-1}\text{ K}^{-1}$ at 100 K) exceed those along $a$-axis (blue, 10.49 $\text{W m}^{-1}\text{ K}^{-1}$) and $b$-axis (red, 10.81 $\text{W m}^{-1}\text{ K}^{-1}$). 

This anisotropy stems from octahedral tilting differentially affecting phonon group velocities and scattering. $\kappa_L$ decreases with temperature increasing (e.g., 2.25 $\text{W m}^{-1}\text{ K}^{-1}$ along $c$-axis at 1000 K) due to intensified phonon-phonon scattering \cite{wang2023role,Hao2024Machine}. The tetragonal phase shows identical $\kappa_L$ along $a$- and $b$-axes (orange, 1.80 to 1.63 $\text{W m}^{-1}\text{ K}^{-1}$ at 1000\,K to 1400\,K) but significantly higher $c$-axis values (around 4.0 $\text{W m}^{-1}\text{ K}^{-1}$). 
The cubic phase exhibits isotropic $\kappa_L$ (around 2 $\text{W m}^{-1}\text{ K}^{-1}$ at 1400 K to 1600 K), reflecting symmetry-minimized anharmonic scattering and positioning it between $a$- and $c$-axes in tetragonal phase.

\begin{figure*}
\centering
\includegraphics[width=0.9\textwidth]{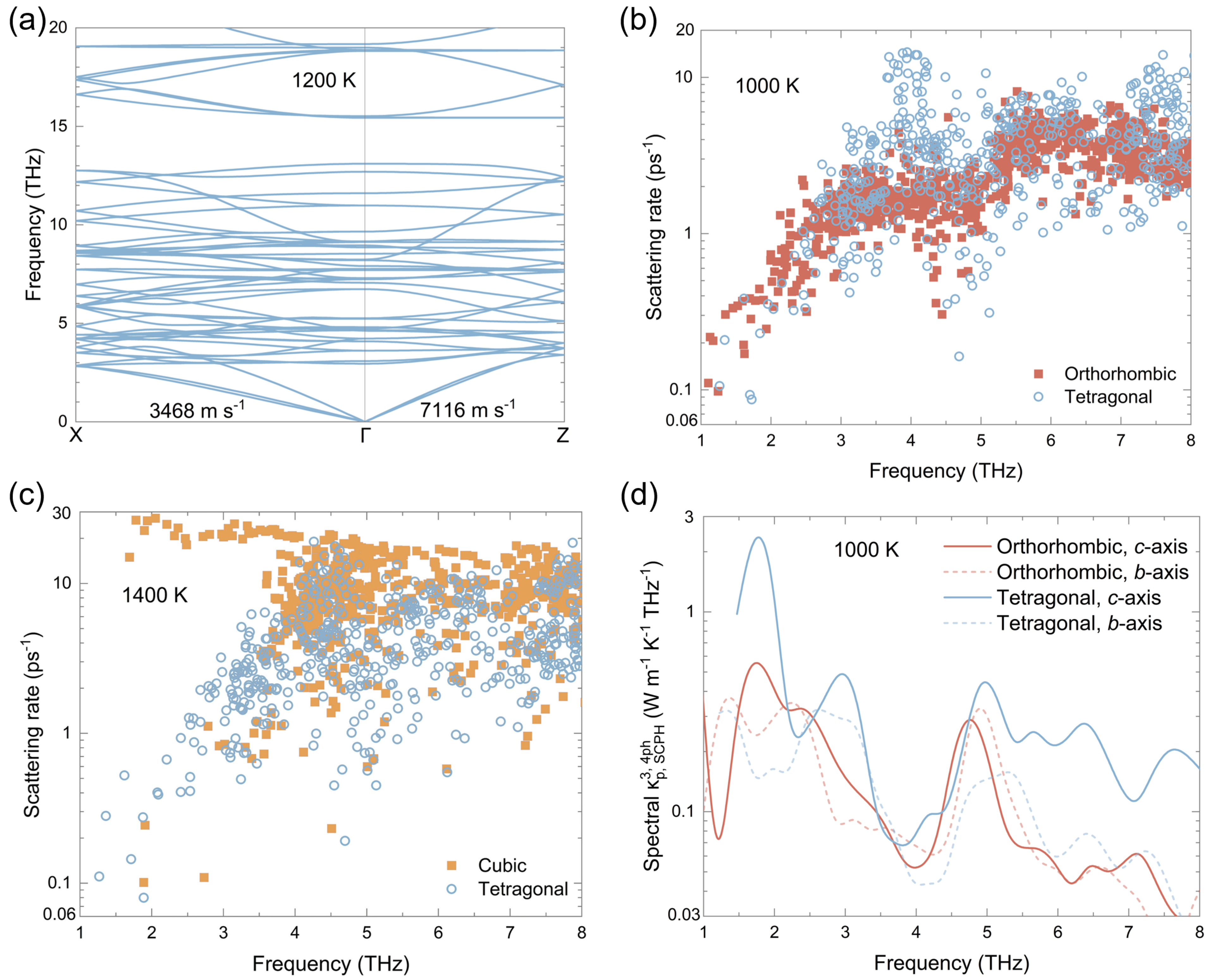}
\caption{(a) Phonon dispersion curves at 1200 K along high-symmetry directions (X, $\Gamma$, Z) in the tetragonal phase.
    (b) Phonon scattering rates as a function of frequency at 1000 K, comparing orthorhombic (red squares) and tetragonal (blue circles) phases. 
    (c) Phonon scattering rates as a function of frequency at 1400 K, comparing cubic (purple squares) and tetragonal (blue circles) phases. 
    (d) Spectral $\kappa_{p, SCPH}^{3, 4ph}$ at 1000 K for different structural phases and orientations: orthorhombic along the $c$-axis (solid red line), orthorhombic along the $b$-axis (dashed red line), tetragonal along the $c$-axis (solid blue line), and tetragonal along the $b$-axis (dashed blue line).
\label{fig3}}
\end{figure*}


The orthorhombic phase demonstrates pronounced anisotropy in Fig.~\ref{fig2}(d). At 300~K, the values of $\kappa_L$ along $a$ and $b$-axes are 80\% and 82\% of that along $c$, respectively, changing to 89\% and 81\% at 1000~K. The tetragonal phase displays even stronger anisotropy, with identical $\kappa_L$ along the $a$- and $b$-axes but only 40\% of that along the $c$-axis. This can be explained by the elevated phonon group velocities along the $c$-axis as evidenced by the steeper slope of the acoustic branch (Fig.~\ref{fig3}(a)) in the $c$-axis directions ($\Gamma$-Z) compared to the $a$,$b$-axes ($\Gamma$--X) in the tetragonal phase SSO at 1200 K. The \(\Gamma\)--Z direction (\(c\)-axis) exhibits significantly steeper acoustic branch slopes than \(\Gamma\)--X (\(a\)/\(b\)-axes), indicating higher phonon group velocities along \(c\)-axis. Quantitatively, group velocity along \(\Gamma\)--Z is approximately 7116 m s\textsuperscript{-1}, more than double the 3468 m s\textsuperscript{-1} along \(\Gamma\)--X.
Enhanced phonon propagation efficiency along \(c\)-axis arises from steeper acoustic branches, enabling longer phonon travel with fewer scattering events.

Fig.~\ref{fig3}(b) compares phonon scattering rates versus frequency (0-8 THz) at 1000 K between orthorhombic (red squares) and tetragonal (blue circles) phases. The Tetragonal phase exhibits similar scattering rates across most frequencies, with a pronounced minimum at around 1.75 THz nearly an order of magnitude lower than orthorhombic phase at same frequency resulting in suppressed phonon-phonon interactions. This suppressed scattering enhances phonon mean free paths (MFPs), contributing to higher \(c\)-axis thermal conductivity in tetragonal phase (4.07 $\text{W m}^{-1}\text{ K}^{-1}$) versus orthorhombic phase (2.25 $\text{W m}^{-1}\text{ K}^{-1}$) at 1000 K. Fig.~\ref{fig3}(d) further analyzes frequency-resolved \(\kappa_p\) at 1000 K, with solid/dashed lines representing \(c\)-/\(b\)-axes contributions and red/blue colors denoting orthorhombic/tetragonal phases. 
To clarify the directional dependence of thermal conductivity, the group velocity of tetragonal SSO is shown in Fig. S6~\cite{SupplementalMaterial}. The group velocity along the \(c\)-axis is higher than that in \(b\)-axis. This pronounced directional advantage in group velocity, coupled with the low scalar scattering rate within low frequency range, accounts for the enhanced thermal conductivity along the \(c\)-axis of tetragonal SSO.

The Tetragonal phase exhibits distinct \(\kappa_p\) peaks at 1.75 THz and 3 THz, corresponding to scattering minima in Fig.~\ref{fig3}(b). These peaks are significantly higher than the orthorhombic phase, indicating phonon transparency windows where weak phonon scattering enables exceptionally large MFPs. Along \(c\)-axis, tetragonal \(\kappa_p\) values are approximately an order of magnitude higher than orthorhombic phase at these frequencies, highlighting low-frequency phonons' critical role. Orthorhombic phase shows broader, less pronounced \(\kappa_p\) distribution without distinct peaks, reflecting higher scattering rates and shorter MFPs. Anisotropy is evident through higher \(c\)-axis contributions in both phases, with tetragonal phase exhibiting stronger directional dependence.

Fig.~\ref{fig3}(c) extends this comparison to 1400 K between cubic (purple squares) and tetragonal (blue circles) phases. Cubic phase displays significantly higher scattering rates across frequency range. Elevated scattering in cubic phase suppresses \(\kappa_L\), yielding lower \(c\)-axis value (\(\kappa_{L,c}^{\mathrm{C}} \approx 2.2\,\text{W m}^{-1}\text{ K}^{-1}\)) versus tetragonal phase (\(\kappa_{L,c}^{\mathrm{T}} \approx 4.0\,\text{W m}^{-1}\text{ K}^{-1}\)) at 1400 K. Reduced scattering in tetragonal phase creates phonon transparency windows facilitating longer MFPs and more efficient thermal transport.

\begin{figure*}
\centering
\includegraphics[width=0.9\textwidth]{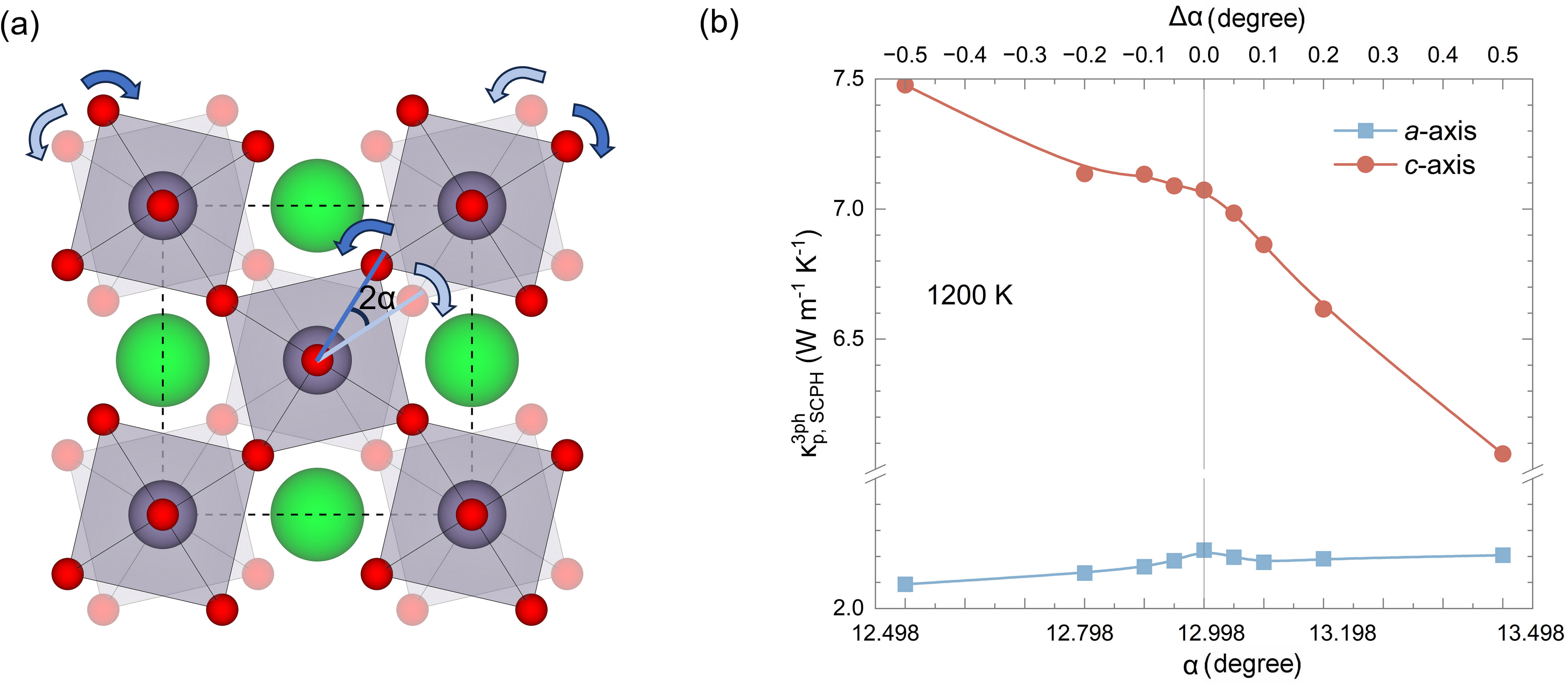}
\caption{(a) Crystal structure of the perovskite material at 1200 K, illustrating the arrangement of atoms in a tetragonal phase with tilting angle of $\alpha = 12.998^\circ$ and blue arrows denoting the rotational directions of the octahedron. The structure shows the unit cell with red spheres representing O atoms, green spheres representing Sr atoms, and gray spheres representing the Sn atoms. (b) $\kappa_{p, SCPH}^{3ph}$ along the $a$-axis and $c$-axis as a function of the angular deviation $\Delta\alpha$ (in degrees) at 1200 K. 
\label{fig4}}
\end{figure*}


Fig.~\ref{fig4}(a) depicts the crystal structure of tetragonal SrSnO\textsubscript{3} (space group $I4/mcm$), showing Sr atoms as green spheres, Sn atoms at centers of gray SnO\textsubscript{6} octahedra, and oxygen (O) atoms as red spheres. The key structural feature is octahedral tilting quantified by angle $\alpha = 12.998^\circ$, with relative rotation between adjacent octahedra along $a$-axis denoted as $2\alpha$. Blue arrows illustrate cooperative rotational directions of SnO\textsubscript{6} octahedra. This tilting originates from a phase transition from cubic phase ($Pm\bar{3}m$, stable above 1295 K) to tetragonal phase (1062-1295 K).

In SrSnO\textsubscript{3}, tilting responds to ionic size mismatch between Sr\textsuperscript{2+} and Sn\textsuperscript{4+}, inducing rotational instability that reduces lattice symmetry. The resulting distortion alters Sn--O bond lengths and angles, directly impacting lattice dynamics and thermal transport properties. Fig.~\ref{fig4}(b) presents $\kappa_L$ along $a$- and $c$-axes versus tilting angular deviation $\Delta\alpha$ (degrees) at 1200 K, with $\alpha = 12.998^\circ$ as the calculated equilibrium tilting angle at this temperature as reference. 

This temperature in Fig.~\ref{fig4} is selected because it falls around the center within the stability range of the tetragonal phase (1062 K to 1295 K), allowing for a focused examination of tilting-induced effects on phonon dispersion and thermal conductivity in this phase. Calculations employ self-consistent phonon (SCPH) approach incorporating 3ph interactions using machine learning potentials (neuroevolution potentials, NEPs). Results reveal pronounced anisotropy: along $c$-axis, $\kappa_L$ decreases significantly from 7.48 W m\textsuperscript{-1} K\textsuperscript{-1} at $\Delta\alpha = -0.5^\circ$ to 6.06 W m\textsuperscript{-1} K\textsuperscript{-1} at $\Delta\alpha = +0.5^\circ$, indicating strong sensitivity to tilting angle variations. Conversely, $a$-axis $\kappa_L$ remains stable at $\sim$2.0 W m\textsuperscript{-1} K\textsuperscript{-1} across same $\Delta\alpha$ range, suggesting weaker tilting dependence. In Fig.~\ref{fig4}(b), only $\kappa_p$ is considered, because $\kappa_c$ is relatively insensitive to the small variations in tilting angle as shown in Fig. S10.

Such tunable anisotropy, where tilting selectively suppresses out-of-plane $\kappa_L$ while preserving in-plane transport, addresses critical demands in directional heat flow control for multilayer devices, enabling efficient thermal management in high-power electronic and thermoelectric applications. This anisotropic behavior stems from tilting-induced modification of phonon dispersion and scattering rates, particularly acoustic mode softening along $c$-axis.

\begin{figure*}
\centering
\includegraphics[width=0.9\textwidth]{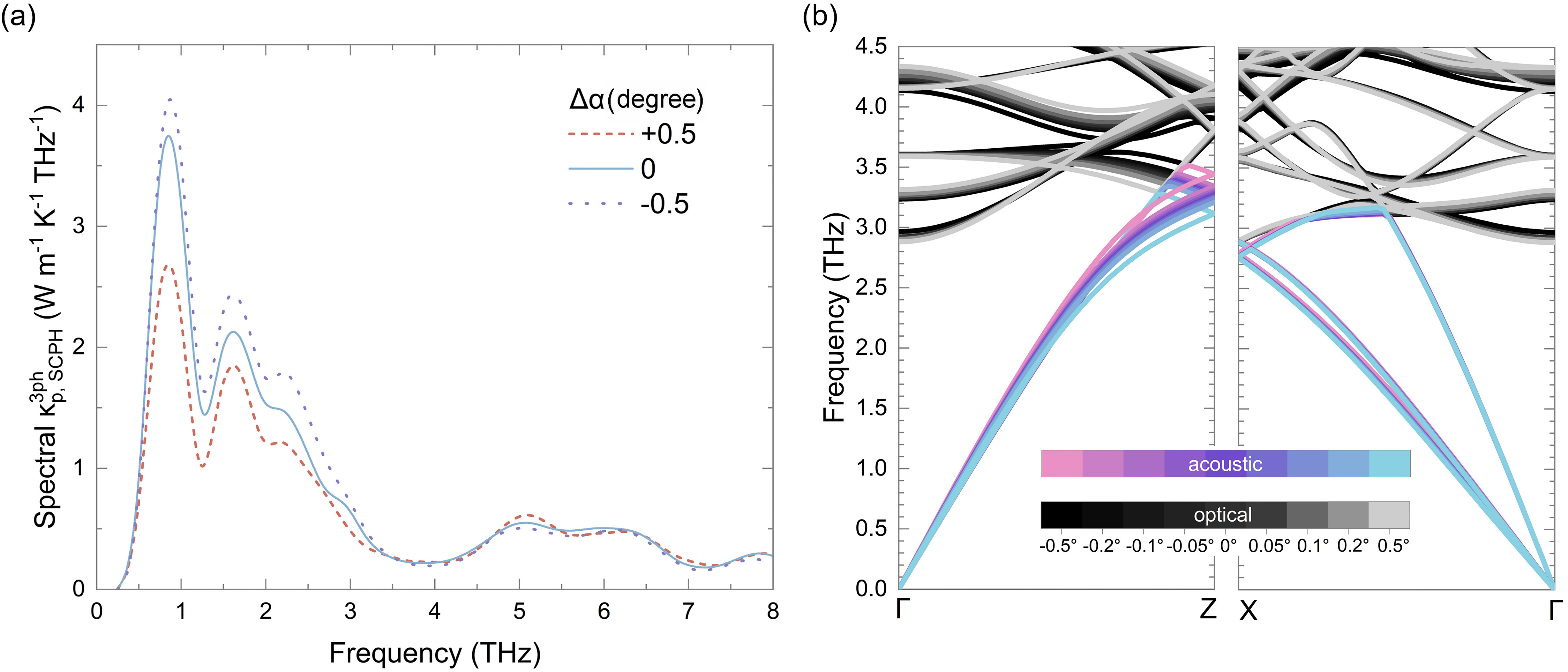}
\caption{(a)Spectral $\kappa_{p, SCPH}^{3ph}$ at 1200 K for tetragonal phase and $c$-axis orientations with $\Delta\alpha$ from $-0.5^\circ$ to $0.5^\circ$.
(b) Tilting angle dependent phonon dispersion curves along high-symmetry paths in the Brillouin zone ($\Gamma$, Z and X), with acoustic (purple to blue) and optical (gray) branches distinguished. 
\label{fig5}}
\end{figure*}

Fig.~\ref{fig5}(a) shows the spectral thermal conductivity ($\partial \kappa_L / \partial f$) as a function of tilting angle. It can be directly observed that their $\kappa_L$ shows a significant difference only below 3 THz. In contrast, comparing with the phonon spectra along high-symmetry paths ($\Gamma$-X and $\Gamma$-Z) in Fig.~\ref{fig5}(b) as a function of tilting angle. It can be seen that below 3 THz, $\kappa_L$ is mainly contributed by acoustic phonons. Along $\Gamma$-Z ($c$-axis direction), acoustic branches soften with increasing tilting angle, evidenced by reduced slopes and lower group velocities. Meanwhile, along $\Gamma$-X ($a$-axis direction), acoustic branches exhibit much slighter change as tilting increases. This is also consistent with the stable $\kappa_L$ of $a$-axis in Fig.~\ref{fig4}(b). 

These suggest that acoustic phonons soften with increasing tilting angle, and the reduction in acoustic phonon group velocity is the primary reason for the decrease in $\kappa_L$ along the $c$-axis direction. Intriguingly, this reduction contrasts with observations in SrTiO$_3$ (STO), where octahedral tilting in the low-temperature tetragonal phase counterintuitively enhances $\kappa_L$ by $\sim$10\% to 20\% relative to the cubic phase, as revealed by \textit{Ab initio} molecular dynamics~\cite{fumega2020understanding}. 

The divergent effects between STO and SSO stem from fundamental differences in lattice anharmonicity and phonon mode coupling driven by the B-site cation: the larger Sn$^{4+}$ ionic radius (0.690 \AA) in SSO promotes stronger chemical bonding asymmetry and lattice softness compared to Ti$^{4+}$ (0.605 \AA) in STO, amplifying tilting-induced acoustic phonon softening in SSO with lowering frequencies and group velocities below 3 THz along the $c$-axis.

\begin{figure*}
\centering
\includegraphics[width=0.9\textwidth]{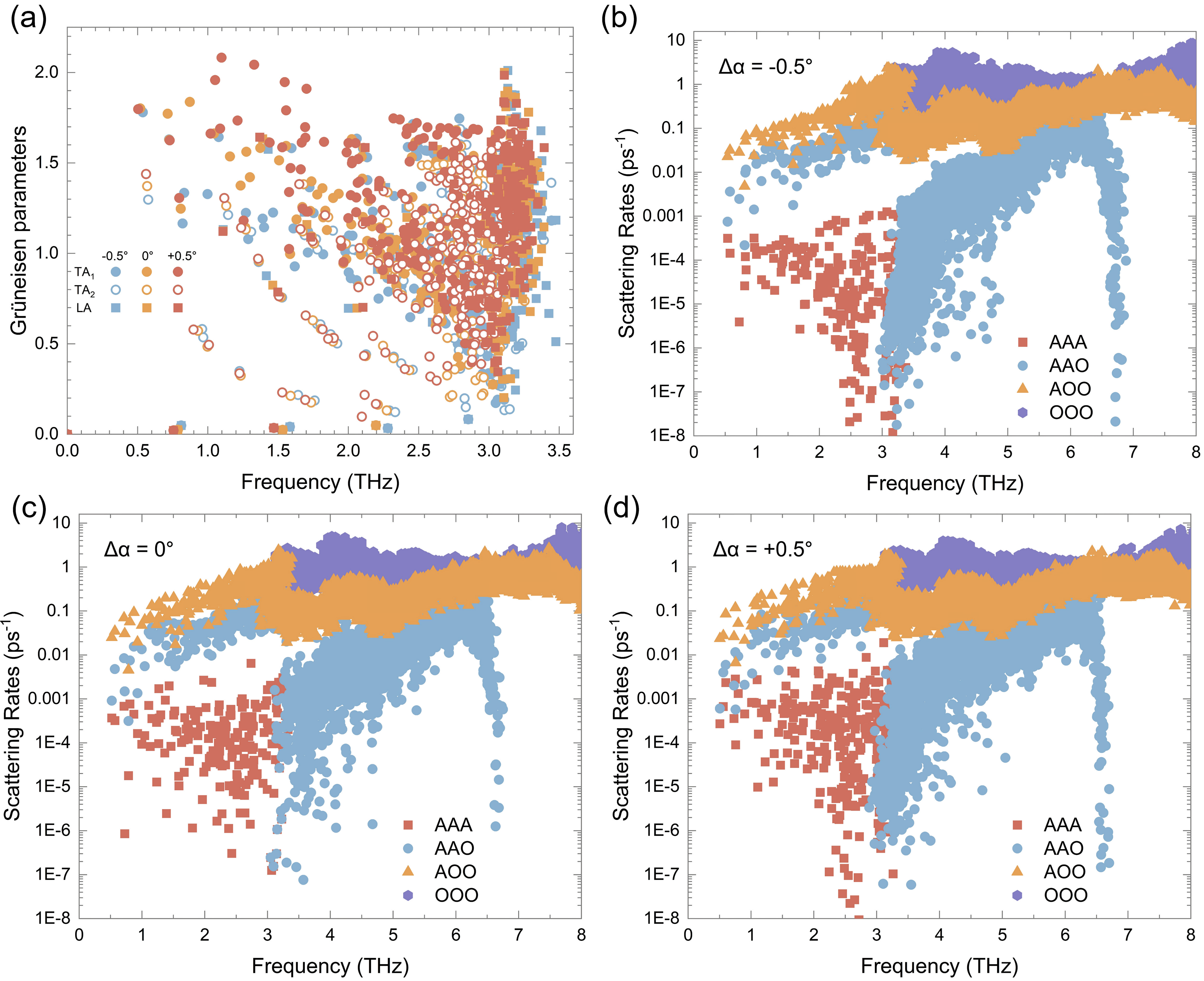}
\caption{(a) Grüneisen parameters as a function of frequency. Phonon scattering rates with (b) $\Delta\alpha = -0.5^\circ$, (c) $\Delta\alpha = 0^\circ$, (d) $\Delta\alpha = +0.5^\circ$, with contributions from AAA (three acoustic), AAO (two acoustic, one optical), AOO (one acoustic, two optical), and OOO (three optical) modes, respectively. 
\label{fig6}}
\end{figure*}

In addition to this, to establish microscopic mechanisms of the reduction in thermal conductivity induced by tilting in Fig.~\ref{fig4}(b), we also systematically analyze the Grüneisen parameters and phonon scattering processes. Fig.~\ref{fig6}(a) plots Grüneisen parameters versus phonon frequency, revealing anharmonic behavior of lattice modes. These parameters, which quantify the volume dependence of the phonon frequencies, increase with the tilting angle, particularly for low-frequency acoustic modes below 2 THz. 

At $\Delta\alpha = +0.5^\circ$ (red), Grüneisen parameters are significantly higher than at $\Delta\alpha = 0^\circ$ (orange) and $\Delta\alpha = -0.5^\circ$ (blue), indicating enhanced anharmonicity with larger tilting angle. This trend reflects amplified nonlinear atomic interactions due to structural distortion, directly contributing to reduced $\kappa_L$ in $c$-axis by increasing phonon scattering.

Fig.~\ref{fig6}(b-d) present phonon scattering rates versus frequency for $\Delta\alpha = -0.5^\circ$, $0^\circ$, and $+0.5^\circ$, respectively, with contributions from scattering channels: AAA (three acoustic), AAO (two acoustic, one optical), AOO (one acoustic, two optical), and OOO (three optical). AAA scattering dominate below 3 THz with much lower scattering rates, while AAO/AOO contributing at higher frequencies. From $\Delta\alpha = -0.5^\circ$ in Fig.~\ref{fig6}(b) to $0^\circ$ in Fig.~\ref{fig6}(c) and $+0.5^\circ$ in Fig.~\ref{fig6}(d), AAA scattering rate rise from around $10^{-4}$ s\textsuperscript{-1} to $10^{-3}$ s\textsuperscript{-1} showing a significant enhancement driven by tilting angle. This mirrors observations where AAA scattering intensifies with structural distortion, shortening phonon lifetimes and directly suppressing $c$-axis $\kappa_L$ shown in Fig.~\ref{fig4}(b).


\section{Conclusion}\label{sec4} 
In this work, we have demonstrated that octahedral tilting in tetragonal SrSnO$_3$ significantly reduces its $\kappa_L$ through the mechanism of acoustic phonon softening. Using DFT calculations augmented by machine learning potentials, we have shown that the tilting-induced distortion leads to a decrease in the frequencies of acoustic phonons, thereby enhancing phonon-phonon interactions, especially AAA processes scattering rates. This effect, the opposite of well-known SrTiO$_3$, is anomalously pronounced along the $c$-axis of SrSnO$_3$, resulting in anisotropic thermal transport properties. Our findings not only provide a detailed understanding of the thermal transport mechanisms in SrSnO$_3$ but also highlight the potential of machine learning approaches in tackling complex phonon dynamics in functional perovskite oxides. These insights could be instrumental in designing materials with tailored thermal properties for applications in thermoelectric devices and thermal management systems.

\section*{Acknowledgements}\label{secAcknow}
The authors gratefully acknowledge discussions with Prof. Xiaojia Wang and Prof. Tianli Feng about the effect of tilting angle on thermal conductivity. 
%
%
Z.G. acknowledge the support from the National Natural Science Foundation of China 
(No.52250191 and No.52595632) and the Fundamental Research Funds for the Central Universities. 
This work is sponsored by the Key Research and Development Program of the Ministry of Science and Technology (No.2023YFB4604100).
We also acknowledge the support by HPC Platform, Xi’an Jiaotong University.

\section*{Author contributions}\label{secContri}
Z.G. designed the simulations and the framework of this research. Y.H. carried out all calculations. 
Y.H. and Z.G. wrote the manuscript. All authors performed data analysis and provided many suggestions. 



\section*{Data Availability}\label{secData}
The structure data of SrSnO$_3$ that we used can be found in Supplemental Material~\cite{SupplementalMaterial}. Other data is available from the authors upon reasonable request.

%

\end{document}